\documentclass{aa}
\usepackage{graphicx}
\newcommand{\be}{\begin{equation}}
\newcommand{\ee}{\end{equation}}
\begin{document}

\title{Radial $B-V/V-K$ color gradients, extinction-free  \\
$Q_{\rm BVK}$ combined color indices,  and the history  \\
of star formation of the Cartwheel ring galaxy}

\author{E.I. Vorobyov\inst{1} \and D. Bizyaev\inst{2}}

\offprints{E.I. Vorobyov}

\institute{Institute of Physics, Stachki 194, Rostov-on-Don, Russia \\
\email{eduard\_vorobev@mail.ru}
\and
Sternberg Astronomical Institute, Universitetskiy prospect 13, Moscow, Russia \\ 
\email{dmbiz@sai.msu.ru}
}

\date{}

\abstract{
In this paper we model and analyse the $B-V/V-K$ radial color gradients observed in the Cartwheel ring galaxy. 
Along with the color-color diagrams, we use the $Q_{\rm BVK}$ combined color indices, which 
minimise the uncertainties in the  observed   $B-V$ and $V-K$  colors introduced by dust extinction.
To model the optical and near-infrared color properties of the Cartwheel galaxy, we assume that
an intruder galaxy generates an expanding ring density wave in the Cartwheel's disk,
which in its turn triggers massive star formation  along the wave's perimeter according to the Schmidt law. 
We use the population synthesis to calculate the color properties of stellar populations
formed in the expanding density wave.
The results of color modelling suggest that the pre-collision Cartwheel was a  
late-type spiral, embedded
in an extensive gaseous disk of sub-critical surface density. The properties of the old stellar disk
are typical for the late-type Freeman disks, with the central surface brightness in {\it V}-band and 
the scale length being $\mu_{\mathrm{V}}^0=21.0 \;\rm mag \; arcsec^{-2}$ and 
$R_0=2$~kpc respectively. The pre-collision gaseous disk has a metallicity gradient 
ranging from $z=z_{\odot}/5$ at the outer regions to $z=z_{\odot}$ in the central regions. 
At present, the  wave of star formation has passed the initial extent of the pre-collision, 
old stellar disk and is currently moving in the predominantly gaseous, 
low-metallicity disk at the radius of 16~kpc. 
Neither young stellar populations formed in an expanding density wave, 
nor their mixture with the old, pre-collision stellar populations can reproduce the 
$B-V$ and $V-K$ colors of the Cartwheel's nucleus+inner ring. 
We find that an additional 10-Myr-old burst of star formation in the nuclear regions, 
along with the visual extinction of $A_{\mathrm{V}}=1^{\mathrm{m}}.3$,  
might be responsible for the peculiar colors of the Cartwheel's nucleus.
\keywords{Galaxies: individual: The Cartwheel -- Galaxies: stellar content -- Galaxies: formation: }
}

\titlerunning{Radial color gradients in the Cartwheel}
\maketitle
        
\section{Introduction}

Ring galaxies are believed to be the result of a head-on galaxy-galaxy collision. 
Such a near-central collision generates an
expanding ring density wave in the disk of a larger galaxy. 
It is likely that star formation will be induced along the wave's perimeter
when the gas density exceeds a threshold (Appleton \& Struck-Marcell \cite{Appleton2}). 
As such a  wave of star formation advances radially from the nucleus, 
it leaves behind evolved stellar populations, with the youngest stars located at the current position of the wave. 
It is expected that this might result in the radial color distribution of stars in the galactic disk, with the 
inner regions being redder than the outer parts of the disk. 
Indeed, observations of the Cartwheel ring galaxy (Marcum et al. \cite{Marcum})
reveal the presence of the optical and near-infrared 
$B-V/V-K$ radial color gradients in the galactic disk. 
Recent observations of a sample of northern ring galaxies by Appleton $\&$ Marston (\cite{Appleton1}) show 
that most of them exhibit radial color gradients 
similar to the pattern observed in the Cartwheel.

Numerical modelling of the Cartwheel's radial color gradients
by Korchagin et al. (\cite{Korch}) shows that the young stellar populations formed in the 
expanding density wave exhibit regular reddening of  colors towards the nucleus only for the sub-solar 
metallicities of the star-forming gas  ($z \le z_{\odot}/2.5$).  Even in the most favorable case of
$z=z_{\odot}/5$, the model colors are much bluer as compared to those observed in the Cartwheel.
Korchagin et al. (\cite{Korch}) argue that the pre-collision disk of old stars is needed to reconcile the
model and the Cartwheel's colors.   
However, their conclusions are based on the direct quantitative 
comparison of the model and the Cartwheel's $B-V$ and $V-K$ colors, 
which  is complicated due to uncertainties in the amount of internal extinction in the Cartwheel's disk. 

In this paper we re-address the question of theoretical modelling of the $B-V/V-K$ 
radial color gradients  observed in the Cartwheel galaxy.
We attempt to exclude the complicated influence of dust extinction on the theoretical interpretation 
of the Cartwheel's $B-V/V-K$ radial color gradients. To do this, we use
insensitive to dust extinction $Q_{\rm BVK}$ combined color indices.  
Such color indices were successfully applied for structure modelling
of dusty galaxies by Zasov $\&$ Moiseev (\cite{Zasov}) and Bizyaev et al. (\cite{Bizyaev}, \cite{Bizyaev2}). 

In Sect.~2 we analyse the $Q_{\rm BVK}$ radial distribution observed in the Cartwheel's disk and
consider the factors defining the value of  $Q_{\rm BVK}$. In Sect.~3 we formulate the adopted
model of star formation in the Cartwheel galaxy. In Sect.~4 we compare 
the model B-V/V-K radial color
gradients and   $Q_{\rm BVK}$ index profiles, obtained in the framework of a first major 
episode of star formation in the Cartwheel's disk, with the observed radial profiles. 
We consider a possibility that the Cartwheel has an old stellar disk typical for the late-type 
spirals. In Sect.~5 we attempt to model the peculiar colors of the Cartwheel's nucleus
and the inner ring. The main results are summarized in Sect.~6.

\section{ Infrared and optical $B-V/V-K$ radial color gradients and $Q_{\rm BVK}$ combined color indices
of the Cartwheel galaxy}

The $B-V/V-K$ radial color gradients of the Cartwheel galaxy are shown in Fig.~1, 
which is a reproduction of Fig.~3 of Marcum et al. (\cite{Marcum}).
The annular points are numbered according to radius, beginning with the inner ring (point~I)
and ending with the outer ring (points~VIII and IX). 
These colors have only been corrected for Galactic reddening using $A_{\rm B}=0^{\mathrm{m}}.22$
and the standard ISM extinction curve (Marcum et al. \cite{Marcum}). 

 Previous  modelling of the Cartwheel's radial color gradient was seriously complicated by the
uncertainties in the amount of internal extinction in the galactic disk.
The internal extinction is only measured towards two bright HII regions 
in the Cartwheel's outer ring ($A_{\rm V}=1^m.94$)
and is not applicable to the whole galactic disk. The value of internal extinction interior to the outer ring 
of $A_{\rm V}=0^{\mathrm{m}}.3$ has been estimated from the available $\mathrm{HI+H_2}$ data (Korchagin et al. \cite{Korch}).
Clearly, correction for the internal extinction  would greatly distort the observed Cartwheel's B-V/V-K radial color pattern,
as shown by correction vectors in Fig.~1, and complicate the task of color interpretation.

\begin{figure}
  \resizebox{\hsize}{!}{\includegraphics{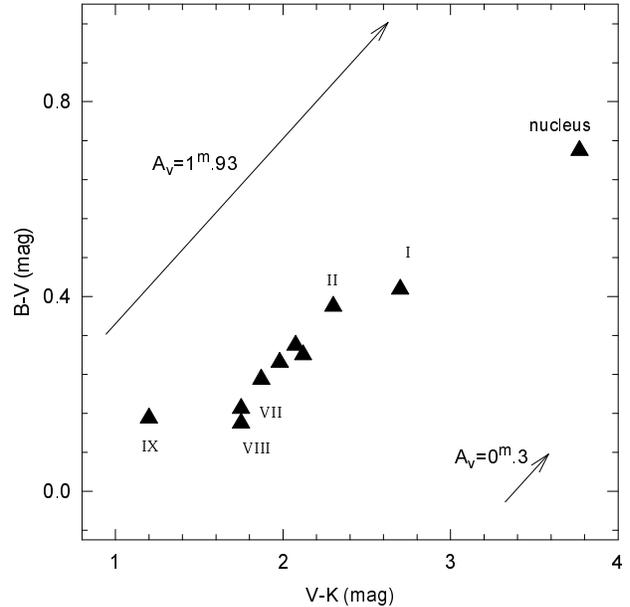}}
      \caption{Radial $B-V/V-K$ color gradients of the Cartwheel galaxy. 
         This  is a reproduction of Fig.~3 
          of Marcum et al. (\cite{Marcum}).
              }
         \label{Fig1}
   \end{figure}

To avoid the complicated influence of dust on the Cartwheel's radial color gradient, we use insensitive to  
dust extinction $Q_{\rm BVK} $ combined color index defined as:
\begin{equation}
Q_{\rm BVK} = (B-V) - {E(B-V) \over E(V-K)} (V-K),
\end{equation}
where $(B-V)$ and $(V-K)$ are colors, and $E(B-V)$ and $E(V-K)$ are color excesses.

For dust properties typical for the Galaxy ($R_{\rm V}=3.1$) the ratio of color excesses $E(B-V)/E(V-K)$ 
is equal to 0.365 (Cardelli et al. \cite{Cardelli}). This value is valid for a "dust shield" approximation, i.e. 
when all dust is located between an observer and a galaxy. For a homogeneous mixture of dust and stars the ratio 
of color excesses $E(B-V)/E(V-K)$
is $10\%$ less than for a "dust shield" approximation (Zasov \& Moiseev \cite{Zasov}) and equal to 0.329. 
As to the Cartwheel galaxy, the internal extinction dominates at the outer ring 
and is comparable to the Galactic reddening in the inner part (Korchagin et al. \cite{Korch}).
Hence, we choose a "homogeneous mixture " approximation rather than a "dust shield" for modelling  
the Cartwheel's color properties.

The following factors define the value of $Q_{\rm BVK}$:

a) contribution of young, blue stars to the integrated flux, i.e. {\it the history of star formation in a certain region of a galaxy}. 
Figures~2a and 2b show the $Q_{\rm BVK}$ combined color indices  of a single starburst superimposed 
on the old stellar population of 10~Gyr. Calculations are performed using the program of Worthey (\cite{Worthey}).
Metallicities of both the young and old stellar populations are equal to $z_{\odot}/5$ ([Fe/H]=-0.7). 
Figure~2a illustrates the influence of  different relative contributions of young stars on the $Q_{\rm BVK}$
indices for a single 50-Myr-old starburst.
Figure~2b shows the dependence of the $Q_{\rm BVK}$ indices on the age of a starburst. 
Here, both the young and old stellar populations
contribute equally to the total stellar mass. 
As it can be seen, $Q_{\rm BVK}$ indices are sensitive to the age of a starburst and
to the relative contribution of young stars.

\begin{figure}
  \resizebox{\hsize}{!}{\includegraphics{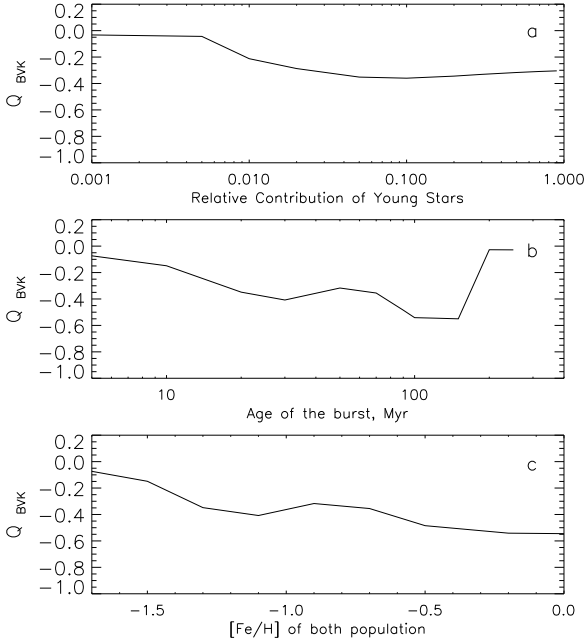}}
      \caption{The $Q_{\rm BVK}$ indices  of a single starburst superimposed 
on the old stellar population of 10~Gyr. {\bf a}) The influence of  different relative contributions of young stars 
on the $Q_{\rm BVK}$ indices for a single starburst of 50~Myr old. 
{\bf b}) Dependence of the $Q_{\rm BVK}$ indices on the age of a starburst.
{\bf c}) The $Q_{\rm BVK}$ indices as a function of metallicities 
of old and young stellar populations (equal for both populations).
              }
         \label{Fig2}
   \end{figure}

Thus, a young burst of star formation should essentially influence the value of $Q_{\rm BVK}$. 
Relative increase of young stars
in a certain region of a galaxy makes its colors bluer, thus, lowering the value of $Q_{\rm BVK}$ (Fig.~2a).
On the other hand, the red supergiant flash (Charlot \& Bruzual \cite{Charlott}) makes the luminosity 
of a young stellar cluster in 
the {\it K}-band to peak at 10~Myr after the burst, which noticeably lowers the value of $Q_{\rm BVK}$ (Fig.~2b).

b) {\it heavy element abundances}. 
Figure~2c gives $Q_{\rm BVK}$ indices as a function of metallicities 
of the old and young stellar populations (equal for both populations). The age of a single starburst is equal to 50~Myr.
Metallicity gradient alone can essentially influence the value of $Q_{\rm BVK}$, 
while other starburst parameters remain unchanged.

c) {\it dust properties}.
Cardelli et al. (\cite{Cardelli}) found that for the dust properties different from those of the Galaxy, i.e. for $R_{\rm V}$ other than 3.1, the
ratio of color excesses $E(B-V)/E(V-K)$ changes. However, the Cartwheel's $Q_{\rm BVK}$ distribution remains qualitatively
unchanged for a fairly wide range of $R_{\rm V}$. Thus, the value of $R_{\rm V}=3.1$ is adopted for the rest of the paper.

Figure~3 shows the Cartwheel's $Q_{\rm BVK}$ distribution as a function of  photometric annulus. Combined color indices are calculated 
using observational data of Marcum et al. (\cite{Marcum}). The numbering of photometric annuli is in accordance with Fig.~3 of Marcum et al. (\cite{Marcum}).
The Cartwheel's $Q_{\rm BVK}$ radial distribution remarkably separates into three sub-profiles, which
correspond to  three structurally different parts of the Cartwheel: annuli~0 and 1 correspond to the nucleus and the inner ring,
annuli~2-7 are the inter-ring region, and annuli~8 and 9 correspond to the outer ring. 
Combined color indices are insensitive to dust extinction and
can be directly understood in terms of stellar populations. This natural separation of $Q_{\rm BVK}$ indices implies that
the nucleus+inner ring, inter-ring region and the outer ring of the Cartwheel consist of  different stellar populations.

 {\it Nucleus and the inner ring.} 
The old stellar content of the nucleus is confirmed by the detection of late-type stellar 
absorption spectrum (Fosbury \& Hawarden \cite{Fosbury}). However, recent mid-infrared 
and $\mathrm{H\alpha}$ observations indicate that the old stellar populations may not totally 
dominate the spectrum. ISOCAM images (Charmandaris et al. \cite{Charm}) show 
significant 7~$\mu m$ and 15~$\mu m$ emission in the nucleus and the inner ring, 
which implies a low-level star formation activity. Detection of weak Ha emission from 
the nucleus and inner ring also indicates that there might be some low-level star formation in the  
Cartwheel's central regions (Amram et al. 1998).

{\it The inter-ring region}. As $\mathrm{H\alpha}$ observations indicate, it is almost totally devoid of any star 
formation activity (Amram et al. \cite{Amram}) and is probably populated by density wave born, post-collision stars with an admixture of old, 
pre-collision stars as implied by progressively reddened $B-V/V-K$ colors towards the Cartwheel's nucleus.
The possibility that dust extinction is responsible for this 
reddening of colors will be tested in numerical simulations in Sect.~\ref{purewave}.

\begin{figure}
  \resizebox{\hsize}{!}{\includegraphics{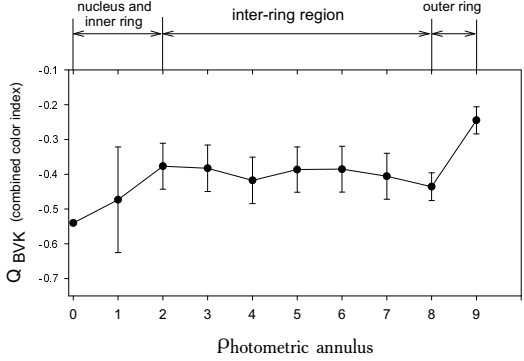}}
      \caption{The Cartwheel's $Q_{\rm BVK}$ distribution as a function of 
photometric annulus.
              }
         \label{Fig3}
   \end{figure}

{\it The outer ring}. Models of ring galaxy formation  predict that the
Cartwheel's outer ring is the current location of an outwardly propagating star formation wave (Appleton \& Struck-Marcell \cite{Appleton2}, 
Struck-Marcell $\&$ Appleton \cite{Struck}). Hence, a large gradient in the $Q_{\rm BVK}$ distribution 
seen across the outer ring in Fig.~3 is indicative of rapid changes 
in the photometric properties of young stellar populations. Probably, the 
outer edge of the ring is so young that it has not yet produced any supergiants, 
whereas the inner edge is in the red-supergiant-flash stage (Marcum et al. 1992).
Existence of the red supergiant flash is confirmed in observations 
of the young stellar cluster NGC~2004 in the LMC (Bica \& Alloin \cite{Bica}).

\section {Theoretical modelling of the Cartwheel's color properties:
orbit crowded density wave plus a Schmidt law of star formation}

To model the optical and near-infrared color properties of the Cartwheel galaxy, we assume that the
head-on galaxy-galaxy collision generates an outwardly propagating  ring density wave in the disk of the Cartwheel.
Additionally, we assume that the collision is central or near-central so that we can neglect the azimuthal variations in
the strength of the density wave. 
Massive star formation in the Cartwheel galaxy is restricted to the outer ring, which is the current location 
of the density wave. Almost no sigh of 
star formation activity is found just behind and ahead of the outer ring (Higdon \cite{Higdon1}).

To imitate the propagating star formation in the Cartwheel's disk,  
we  assume that the star formation rate is proportional to  a ring density enhancement  expanding 
radially in a gaseous disk with velocity $v$. The density enhancement is given by a Gaussian function:
\begin{equation}
C(r-vt)=A(r) \; \exp\left[ - {(r-vt)^2 \over L^2}\right],
\end{equation}
where $A(r)$ is the amplitude of the density enhancement and $L$ is a half-width of the density enhancement.
The balance of surface density of stars ($\Sigma_s$) at a distance $r$ from the center and time $t$ can be then written as follows
(Korchagin et al. \cite{Korch}):
\begin{equation}
{d \Sigma_s \over dt}=-D +  a\; C^n(r-vt),
\label{eq1}
\end{equation}
where the first term in the right-hand side of  Eq. $\ref{eq1}$ is the rate of death of stars:
\begin{equation}
D={1 \over \sum \limits_{m_1}^{m_2} m^{1-\alpha}} \; \sum \limits_{m_1}^{m_2} m^{1-\alpha}\; a\; C^n(r-v(t-\tau_m)),
\end{equation}
where $\tau_m$ is the life-time of a star of mass $m$, and $\alpha$ is the slope of the IMF. 

We consider  a circular density
enhancement propagating from the center of a gaseous disk to the present location of the 
Cartwheel's outer ring at 16~kpc 
$(H_0=100\;  \rm km\; s^{-1}\;  Mpc^{-1})$.
The computational area is divided into 10 annuli as in Plate 1 of Marcum et al. (\cite{Marcum}). 
Annulus~1 is two kiloparsec wide and centered at 3.1~kpc, thus covering the current location of the inner ring. The centroids of annuli
2-9 are separated by 1.25~kpc, with the 2-nd annulus centered at 6.8~kpc. 
The 0-th annulus represents the nucleus.
The mass of stars  formed in each annulus is distributed into individual stellar masses 
using the Salpeter IMF with $\alpha=1.35$
and the stellar mass interval of $m_1$=0.1~$M_{\odot}$ and $m_2$=100~$M_{\odot}$.
The luminosity of each annulus at a given band $A$ after time t can be computed as:
\be
L_A(t)=\sum \limits_m \sum \limits_{\tau{_m}} l_A(m,\tau_m) N(m,\tau_m,t),
\label{eq5}
\ee
where $N(m,\tau_m,t)$ is the number of stars in a given annulus at time $t$ with mass $m$ and age $\tau_m$,
and $l_A(m,\tau_m)$ is the luminosity of a star in the band $A$.
Stellar luminosities $l_A(m,\tau_m)$ are obtained using stellar evolutionary (Schaller et al. \cite{Schaller}) and atmospheric (Kurucz \cite{Kurucz})
models. The population synthesis code used in this work is explained in 
detail in Mayya (\cite{Mayya1}, \cite{Mayya2}). 
This code synthesizes a number of observable quantities in the optical and near-infrared
parts of the spectrum, which are suitable for comparison with the observed 
properties of giant star-forming complexes. Since the Cartwheel is a recent phenomenon
with high rates of star formation, this code is especially suitable for our purposes.
The results of this code are compared with those of other existing codes by Charlot (\cite{Charlott2}).

Classic Schmidt law of star formation ($n=1.5$) is assumed throughout the paper (Kennicutt \cite{Kennicutt2}). 
The half-width of a density enhancement $L$ is chosen so as to reproduce the observed $\mathrm{H\alpha}$ 
surface brightness profile in the
Cartwheel's outer ring (Higdon \cite{Higdon1}). This gives the value of $L=1$~kpc. 
Velocity of a propagating density enhancement 
$v=55$~km~s$^{-1}$ is assumed to coincide
with the expansion velocity of the outer ring $v=53 \pm 9\; \rm km \; s^{-1}$, 
the value derived by Higdon (\cite{Higdon2}) from HI kinematics of the outer ring. 

With the values of  $L$ and $v$ being fixed, the coefficient $a$ and the amplitude of density enhancement
$A(r)$ constitute the set of parameters, which determine the number of stars $N(m,\tau_m, t)$ formed by a density wave
in each photometric annulus (see eq. (\ref{eq5})). The birthrate of stars $a$ relates the star formation rate
observed in the Cartwheel galaxy with the assumed  Schmidt law of star formation (see eq. (\ref{eq1})). 
We estimate its value for the Cartwheel's outer ring using the measured rate of star formation 
67~$M_{\odot}\; \rm yr^{-1}$ (Higdon \cite{Higdon1}), the
mean surface density of gas $\Sigma_{\rm gas} \approx 10\; M_{\odot} \;\rm pc^{-2}$ (Higdon \cite{Higdon2}) 
and   the outer ring's area of  250~kpc$^2$.
This gives the dimensionless value of $a=0.0025$  when masses, time scales, and distance scales are expressed in
units $10^7\; M_{\odot}$, $10^6$ yrs, and 1~kpc respectively. 

We assume that the amplitude of density enhancement $A(r)$ is twice as big as the unperturbed, 
pre-collision gas surface density, which is typical for the spirals and is confirmed in
numerical simulations by Appleton $\&$ Struck-Marcell (\cite{Appleton2}).    
To constrain the pre-collision radial profile of gas, we fit the model {\it R-}band and $\mathrm{H\alpha}$  
surface brightness profiles of stellar populations formed in the expanding density wave 
to  the profiles observed in the Cartwheel. We do this by
varying the radial profile of gas  in the pre-collision galactic disk.

\section{Model B-V/V-K radial color gradients and $Q_{\rm BVK}$ indices of the Cartwheel
galaxy.} 
\label{purewave}
\subsection{Density wave in a purely gaseous disk. }
\label{denswave}

As mentioned in the introduction, Korchagin et al. (\cite{Korch}) found that the density 
wave propagating in {\it  a purely gaseous disk of uniform metallicity} cannot reproduce
the Cartwheel's $B-V/V-K$ radial color gradient. The model colors are much bluer 
as compared to those observed in the Cartwheel. Korchagin et al. (\cite{Korch}) argued 
that the pre-collision disk of old stars was needed to reconcile the model and the Cartwheel's colors. 
We mainly confirm their conclusion. However, use of the extinction-free $Q_{\rm BVK}$ 
indices allows us to notice some interesting features that have been missed 
by Korchagin et al. (\cite{Korch}).

As in Korchagin et al. (\cite{Korch}), we consider the expanding density wave in a purely gaseous 
disk with five different metallicities: $z_{\odot}/20$, $z_{\odot}/5$, $z_{\odot}/2.5$,
$z_{\odot}$, and $2 z_{\odot}$. Our computations show that the model $B-V$ and 
$V-K$ colors are bluer as compared to the Cartwheel's colors for each metallicity 
of the star-forming gas.
If dust extinction were solely responsible for this difference between the model and 
observed colors, then the model $Q_{\rm BVK}$ indices would lie within 
the observed limits. Indeed, use of the $Q_{\rm BVK}$
indices minimises the uncertainties in observed colors introduced by dust extinction.
However, we find that the model $Q_{\rm BVK}$ indices lie beyond the observed limits {\it for most of
the photometric annuli}, thus indicating that the reddening of the observed colors with respect to
the model colors {\it of those annuli} cannot be attributed to the extinction 
gradient in the Cartwheel's disk.

In case of $z_{\odot}/2.5$ and $z_{\odot}/5$, however, the $Q_{\rm BVK}$ indices
of the Cartwheel's outer region (annuli VI-IX) 
lie within the observed limits, thus indicating that dust extinction could 
be responsible for the reddening of the observed colors in this region. 
Here, we note that Korchagin et al. (\cite{Korch})
used a uniform metallicity of $z_{\odot}/20$  to model the Cartwheel's radial color gradient.
While this choice is certainly justified for the Cartwheel's outer ring (Fosbury \& Hawarden \cite{Fosbury}), 
it is hard to believe that such a low metallicity is present in the Cartwheel's central regions. 
Due to very weak line emission, there have been no measurements of heavy element 
abundances interior to the outer ring. 
However, it is known that most galactic disks have a metallicity gradient, with metallicities in the
nucleus up to $\sim$ 10 times higher than in the outer parts of the disk 
(Smartt $\&$ Rolleston \cite{Smartt}).
In this paper we assume that metallicity gradient is present in the pre-collision gaseous disk of 
the Cartwheel galaxy and choose its value so as to provide a better fit
of the model $Q_{\rm BVK}$ indices with those observed in the Cartwheel's disk. 
The best fit is found for the metallicity gradient
spanning the range from $z=z_{\odot}/3.75$ at annuli IX, VIII, and VII, $z=z_{\odot}/2.5$ 
at annuli VI and V, to $z=z_{\odot}$ at annuli IV-nucleus.

The open circles in Fig. 4a show the model $B-V/V-K$ colors of stellar populations formed in the
density wave propagating in a purely gaseous disk with the metallicity gradient given above.
The open circles in Fig. 4b illustrate the corresponding model $Q_{\rm BVK}$ indices.
The outer three photometric annuli (9, 8, and 7) and the nucleus are identified both for 
the model colors and the colors observed in the Cartwheel galaxy, 
the latter being shown by the filled triangles with error bars.
The model $B-V/V-K$ colors do not exhibit a regular reddening towards the nucleus and 
are bluer as compared to the Cartwheel's colors.
On the same time, the model and the Cartwheel's  $Q_{\rm BVK}$ indices 
show a good correspondence for the outer ring and two inner annuli adjacent to the outer ring, 
implying that the difference of the corresponding $B-V/V-K$
colors in Fig.~4a might be related to the internal extinction gradient existing in 
the Cartwheel's disk.

\begin{figure}
  \resizebox{\hsize}{!}{\includegraphics{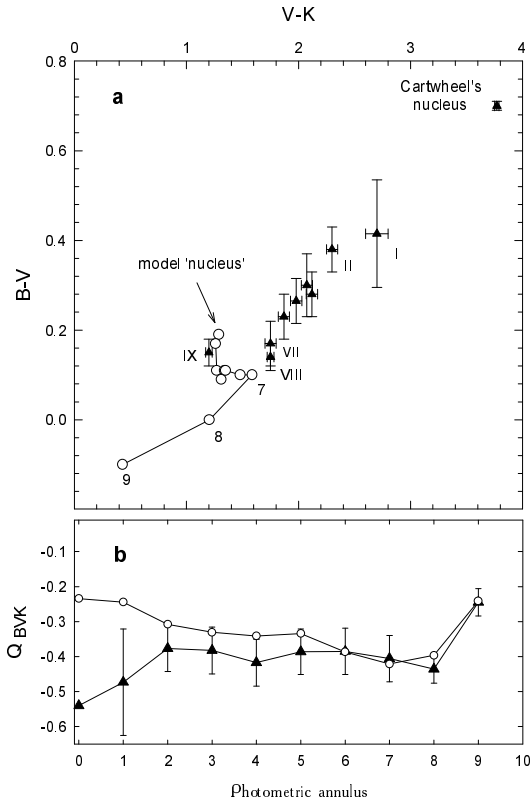}}
 \caption{({\bf a}) The $B-V$ and $V-K$ color-color diagram and ({\bf b}) 
the $Q_{\rm BVK}$ combined index profiles. The open circles show the model profiles 
for the density wave of amplitude $A(r)=70 \: \exp(-r/12)\: M_{\odot}\:\rm pc^{-2}$ 
and velocity $v=55\:\rm km\: s^{-1}$ expanding radially in a purely gaseous disk.
The metallicity gradient of the gaseous disk ranges from $z=z_{\odot}$ at the 
nucleus+annuli I-IV, $z=z_{\odot}/2.5$ at annuli V-VI, to $z=z_{\odot}/3.75$ 
at annuli VII-IX. The filled triangles with error bars show the radial color gradients 
and index profiles observed in the Cartwheel galaxy. 
              }
         \label{Fig4}
   \end{figure}  

Now we consider an extreme case and assume that the difference of the model and observed 
$B-V/V-K$ colors in Fig.~4a  is solely due to  internal extinction gradient in the Cartwheel's disk.
We warn, however, that this assumption is applicable only to those annuli, 
which $Q_{\rm BVK}$ indices lie within the observed limits, 
as in this case their colors are reddened along the correction vector.
The values of color excess $E(B-V)$, implied by this difference, are listed 
in Col.~2  of Table~1. The surface densities of gas $\Sigma_{\rm HI+H_{2}}^{\rm ext}$ 
in Col.~3 are obtained from Col.~2 using the standard mass-to-dust ratio and 
$R_{\rm V}=3.1$ (Bohlin et al. \cite{Bohlin}). Column~4  in Table~1 gives 
the observed surface densities  of atomic and molecular hydrogen 
$\Sigma_{\rm HI+H_{2}}^{\rm obs}$ in each annulus 
(Higdon \cite{Higdon2}, Horellou et al. \cite{Horellou}).

\begin{table}[h]
\caption{\label{Table1}Extinction-estimated and observed surface densities of HI and $\mathrm{H_2}$ in the Cartwheel}
\vskip 0.1cm
\begin{tabular}{llll}
\hline
\hline
annulus & $E(B-V)$  & $\Sigma_{\rm HI+H_{2}}^{\rm ext}$ & 
$\Sigma_{\rm HI+H_{2}}^{\rm obs}$ \\ [2 pt]
  & (mag) &  $M_{\odot} \;\rm pc^{- 2}$  & $M_{\odot}\;\rm pc^{-2}$ \\
\hline
VI & $0^m.13$ & 5.7 & $\le 4.5$  \\ 
VII & $0^m.07$ & 3.1 & $ \le 4.5$  \\ 
VIII & $0^m.14$ & 6.1 & 9.3-16 \\ 
IX & $0^m.25$ & 10.9 & 9.3-16  \\
\hline
\end{tabular}

\end{table}

It can be noticed that the extinction-estimated gas densities in Col.~3 are below the detection upper
limit for the outer ring+annulus~VII. 
Hence, the internal extinction gradient can indeed be responsible for the difference between
the observed and model $B-V/V-K$ colors of this region in Fig.~4a.

Very large values of $\mathrm{H{\alpha}}$ equivalent widths were found by 
Higdon (\cite{Higdon1}) in the Cartwheel's outer ring, especially in the ring's southern quadrant, 
which peaks at $EW_{\mathrm{H{\alpha}}}=1250\; \mathrm{\AA}$ and exceeds
$400\;  \mathrm{\AA}$ everywhere. The $\mathrm{H{\alpha}}$ equivalent width 
($EW_{\mathrm{H{\alpha}}}$) is sensitive to the ratio of ionizing, 
massive ($>10\; M_{\odot}$) stars to lower mass red giant stars and can be used 
to measure the relative strengths of these two populations (Kennicutt \cite{Kennicutt1}).
Table~2 presents  $EW_{\mathrm{H{\alpha}}}$ of stellar populations born in the expanding ring 
density wave. $EW_{\mathrm{H{\alpha}}}$ are computed  at the time when the wave reaches 
the present position of the Cartwheel's outer ring at 16~kpc. All parameters of Fig.~4 are retained.
As it is seen in Table~2, the averaged over
annuli VIII and IX model value of  $335\; \mathrm{\AA}$  is close to
the azimuthally averaged value of $360 \; \mathrm{\AA}$ measured 
in the Cartwheel's outer ring by Higdon (\cite{Higdon1}).
The large model and observed $EW_{\mathrm{H{\alpha}}}$ in the outer ring 
indicate that the Cartwheel is experiencing the first major episode of star formation {\it at this radius}.

\begin{table}[h]
\caption{\label{Table2} Model $\mathrm{H\alpha}$ equivalent widths}
\vskip 0.1 cm
\begin{tabular}{lllll}
\hline
\hline
annulus &  VI & VII & VIII & IX \\ 
$EW_{\mathrm{H{\alpha}}}\; (\rm \AA)$ & 0.03 & 0.3 &53 & 620 \\
\hline
\end{tabular}
\end{table}

Korchagin et al. (\cite{Korch}) argued that the colors of the Cartwheel galaxy cannot be reproduced
by a density wave propagating in a purely gaseous disk irrespective of its metallicity. 
While this is certainly true for the 
Cartwheel's inner region, we find that the observed $B-V/V-K$ colors and $Q_{\rm BVK}$ 
indices of {\it the outer ring+annulus~VII} can be well reproduced by a density wave 
propagating in a purely gaseous disk of $z=z_{\odot}/3.75$. 
 
\subsection {The pre-collision old stellar disk.}
\label{precol}

\begin{figure*}
 \centering
  \includegraphics[width=14 cm]{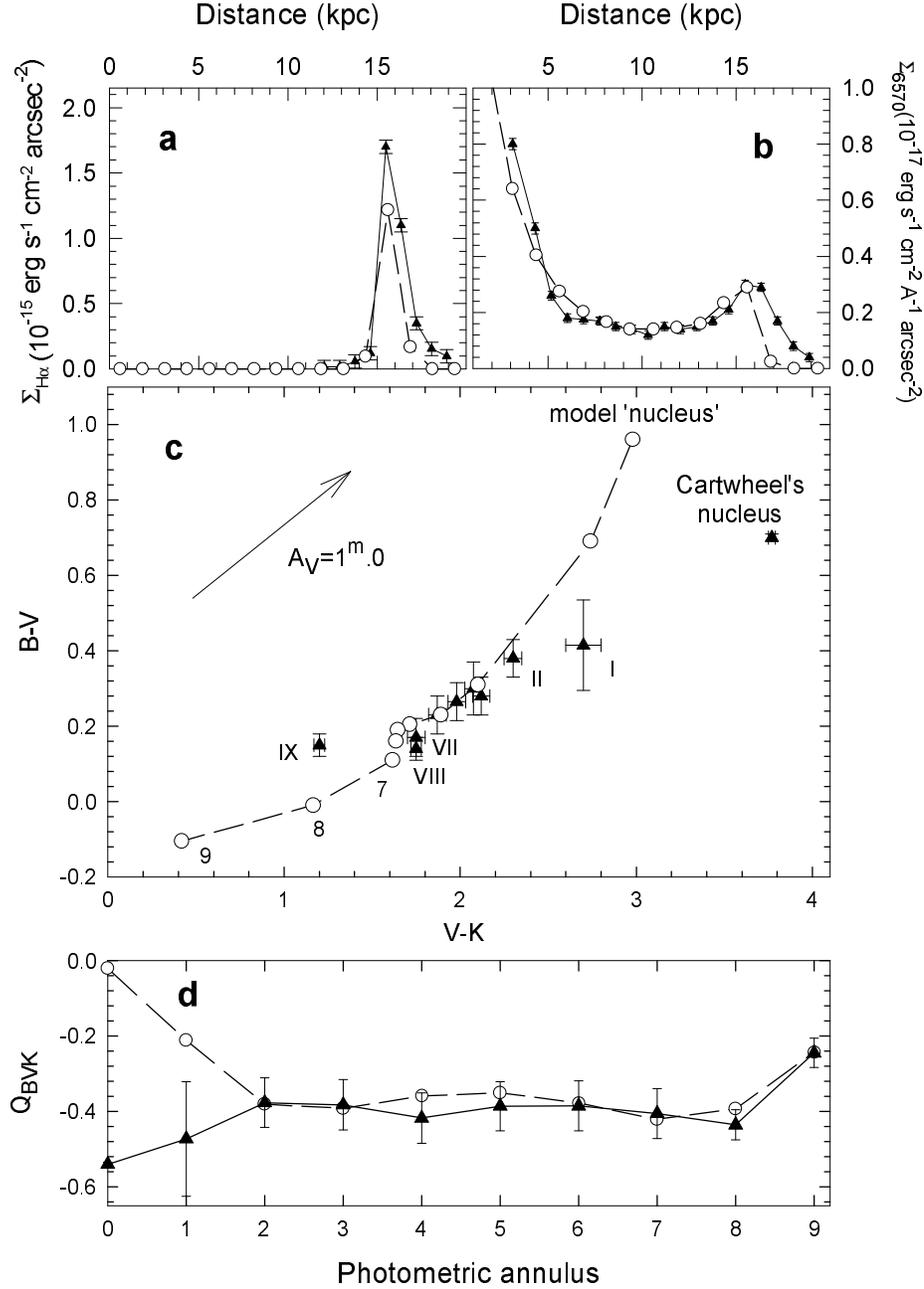}
      \caption{The radial surface brightness profiles in ({\bf a}) $\mathrm{H\alpha}$  and ({\bf b})
{\it R}-band, ({\bf c}) the $B-V/V-K$ color-color diagram, 
and ({\bf d}) the $Q_{\rm BVK}$ radial profiles. 
The open circles show the model profiles for the density wave of amplitude
 $A(r)=50 \: \exp(-r/15)\: M_{\odot}\:\rm pc^{-2}$ and velocity
 $v=55 \:\rm km\: s^{-1}$ expanding radially in the galactic disk.  The central surface brightness 
in the {\it V}-band of the old stellar component of the galactic disk 
 is $\mu_{\mathrm{V}}^0=21.0 \;\rm mag \; arcsec^{-2}$ and the scale length is $R_0=2$~kpc. 
The gaseous component has the metallicity gradient ranging from $z=z_{\odot}$ at
the nucleus+annuli~I-III to $z=z_{\odot}/5$ at annuli~IV-IX. The filled triangles with error bars 
show ({\bf a,b}) the measured radial surface brightness profiles, ({\bf c}) the radial color gradients 
and ({\bf d}) index profiles of the Cartwheel galaxy.}
         \label{Fig5}
   \end{figure*}  

To reconcile the model and observed $B-V/V-K$ colors and $Q_{\rm BVK}$
indices of the Cartwheel's inner region in Fig.~4 (annuli VI-nucleus),
we follow the prescriptions given in Korchagin et al. (\cite{Korch}) and consider
a possibility that the pre-collision Cartwheel had an old stellar disk.
We assume that the properties of the old stellar disk are 
typical for the late-type Freeman disks ($\mu_V^0=21.0 \;\rm mag \; arcsec^{-2}$ 
and $R_0=2$~kpc) rather than for  low surface 
brightness galaxies ($\mu_V^0=23.5 \;\rm mag \; arcsec^{-2}$ 
and $R_0=5$~kpc) as adopted by Korchagin et al. (\cite{Korch}).
Metallicity of the pre-collision gaseous disk  is assumed to span the range from 
$z=z_{\odot}/5$ at six outermost annuli to $z=z_{\odot}$ at the inner annuli and the nucleus. 
Galactic collision generates an expanding density wave, which adds the newly born 
stellar populations to the old stellar populations existed before the collision.

Figures~5a and 5b show the resulting model $\mathrm{H{\alpha}}$ and $R$-band 
surface brightness profiles. Interior to the outer ring, the model $R$-band profile 
shows a good correspondence to the observed profile, as illustrated in Fig.~5b by 
the open circles and filled triangles respectively.
Measured by Higdon (\cite{Higdon1}) $R$-band surface brightness of the inner ring 
is roughly twice the peak value of the outer ring and this feature is well reproduced in Fig.~5b. 

\begin{table}[h]
\caption{\label{Table3} Extinction-estimated and observed surface densities of HI and $\mathrm{H_2}$ in the Cartwheel}
\vskip 0.1 cm
\begin{tabular}{llll}
\hline
\hline
annulus & $E(B-V)$  & $\Sigma_{\rm HI+H_{2}}^{\rm ext}$ & 
$\Sigma_{\rm HI+H_{2}}^{\rm obs}$ \\ [2pt]
  & (mag)  & $M_{\odot}\;\rm pc^{-2}$ & $M_{\odot}\;\rm pc^{-2}$ \\
\hline
II & $0^m.07$ & 3.1 &  $\le 4.5$  \\ 
III & $0^m.07$ & 3.1 & $\le 4.5$  \\ 
IV & $0^m.07$ & 3.1 & $\le 4.5$ \\ 
V & $0^m.07$ & 3.1 & $\le 4.5$  \\ 
VI & $0^m.07$ & 3.1 & $\le 4.5$  \\ 
VII & $0^m.06$ & 2.6 & $ \le 4.5$  \\ 
VIII & $0^m.15$ & 6.6 & 9.3-16 \\ 
IX & $0^m.26$ & 11.4 & 9.3-16  \\
\hline
\end{tabular}
\end{table}

The model radial $B-V/V-K$ colors, shown by the open circles in Fig.~5c, 
exhibit  regular reddening towards the nucleus reminiscent of that observed in the Cartwheel's disk. 
The sequence of $Q_{\rm BVK}$ indices observed in the Cartwheel's outer ring 
and the inter-ring region is well reproduced, as shown by the open circles in Fig.~5d. 
This implies that the difference of the model and observed colors for each photometric 
annulus can be understood in terms of internal extinction. 
This is certainly not true 
for the Cartwheel's inner ring and the nucleus, as their $B-V/V-K$ colors 
are not reddened along the correction vector with respect to the model colors. 
This results in a substantial disagreement of the corresponding $Q_{\rm BVK}$ indices. 
As in Sect.~\ref{denswave}, we consider an extreme case and assume that the 
difference of the model and observed $B-V/V-K$ colors of the outer ring and the inter-ring region
 in  Fig.~5c  is solely due to  internal extinction gradient in the Cartwheel's disk.
The values of $E(B-V)$ implied by this difference are listed 
in Col.~2  of Table~3.

It can be noticed that the extinction-estimated gas densities in Col.~3 are below the detection upper
limit in Col.~4 for  all annuli. Hence, the internal extinction gradient can be responsible for the 
difference between the observed and model $B-V/V-K$ colors of the inter-ring region 
and the outer ring in Fig.~5c.

Comparison of Figs. 4a and  5c shows that the model $B-V/V-K$ colors of 
the outer ring+annulus~VII in Fig. 5c
are not affected noticeably by  the old stellar populations. 
Indeed, Fig.~6 illustrates the ratio of fluxes of the old stellar populations to 
the  young density-wave-born stellar populations in {\it V}-band ($F_{\rm V}[o/dw]$) and
{\it K}-band ($F_{\rm K}[o/dw]$) as a function of galactic radius.   
It is seen that the old stellar populations dominate at the smaller radii ($R \le 3$~kpc) 
in both bands. At the intermediate radii ($3 < R \le 12$~kpc), input from the young 
stellar populations to the total flux in {\it V}-band is  ten times (on average) 
the input from the old stellar populations, while both stellar populations contribute 
comparable fluxes in {\it K}-band. The young stellar populations totally dominate 
at the larger radii ($R > 12$~kpc ) in both bands.
This indicates that the colors of the Cartwheel's outer ring {\it are determined by the young
density-wave-born stellar populations} and can be reproduced by a density wave 
propagating in a purely gaseous disk, as argued in Sect. \ref{denswave}.
We note, however, that the interpretation of colors
of the Cartwheel's inner region do require the presence of the old pre-collision stellar disk, which
is in agreement with conclusions of Korchagin et al. (\cite{Korch}).

\begin{figure}
   \resizebox{\hsize}{!}{\includegraphics{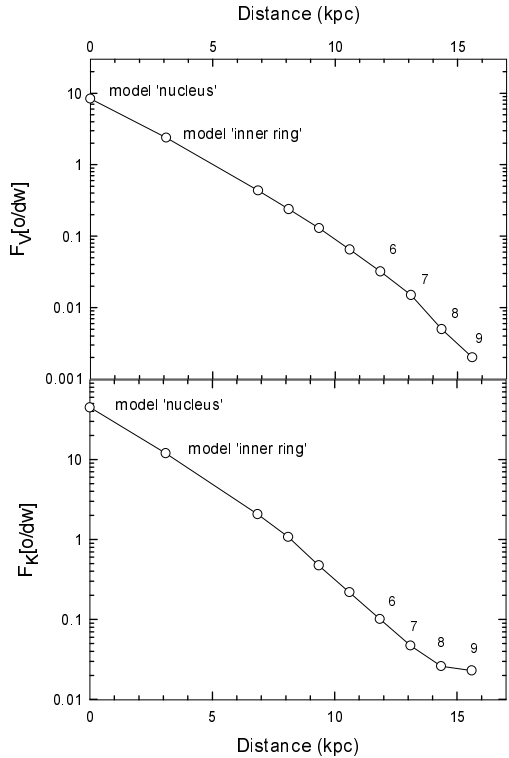}}
      \caption{The ratio of fluxes in {\it V}-band (the upper frame) and {\it K}-band (the lower frame)
of the old, pre-collision stellar populations to the young stellar populations formed
in the density wave. All parameters of 
Fig.~5 are retained. The numbering of annuli is in accordance with Fig. 1.}
         \label{Fig6}
   \end{figure}  

Thus, theoretical modelling of the color gradients observed in the Cartwheel galaxy 
suggests that the pre-collision Cartwheel was a late-type spiral ($R<12$~kpc)  
embedded into an extensive HI disk. The subcritical gas disk of 
$\Sigma_{\rm HI} \le 11\; M_{\odot} \;\rm pc^{-2}$ (Higdon \cite{Higdon2})
at larger radii ($R \ge 12$~kpc) prevented robust massive star formation (MSF) 
and chemical enrichment over the disk's lifetime. 
The passage of a companion galaxy initiated an outwardly propagating ring density wave, 
which triggered high rates of MSF along it's perimeter.
The  density wave has passed the original extend of the pre-collision
stellar disk and  is currently moving in the predominantly gaseous, 
low-metallicity disk at the radius of the Cartwheel's outer ring ($R \approx 16$~kpc).

Another result of color modelling is a possible presence of a 
 large internal extinction gradient across the outer ring of the Cartwheel (annuli VIII and IX). 
Indeed, if the difference between the model and observed $B-V/V-K$ colors of the  
Cartwheel's outer ring in Figs.~4a and
5c can be attributed to the internal extinction as shown above, then  $E(B-V)$ 
in the outer edge of the outer ring (annulus IX)
is about two times
higher as compared to $E(B-V)$ in the inner edge of the outer ring (annulus VIII).
Models of ring galaxies  predict that the Cartwheel's outer ring is the current 
location of a star formation wave (Appleton $\&$ Struck-Marcell  \cite{Appleton2}, 
Struck-Marcell $\&$ Appleton \cite{Struck}). 
The star formation wave is expected to compress the gas ahead by a coherent action 
of shock waves, expanding superbubbles, ultraviolet radiation from massive stars, ets., 
thus increasing the value of extinction at the leading edge of the wave. 
Observationally proved presence of such an extinction gradient across
the outer ring might be a powerful confirmation of propagating nature of star formation 
in ring galaxies. More sophisticated models of star formation in the ring galaxies are  
needed to  investigate these aspects.

\section{The inner ring and the nucleus}
\label{innerring}

\begin{figure}
  \resizebox{\hsize}{!}{\includegraphics{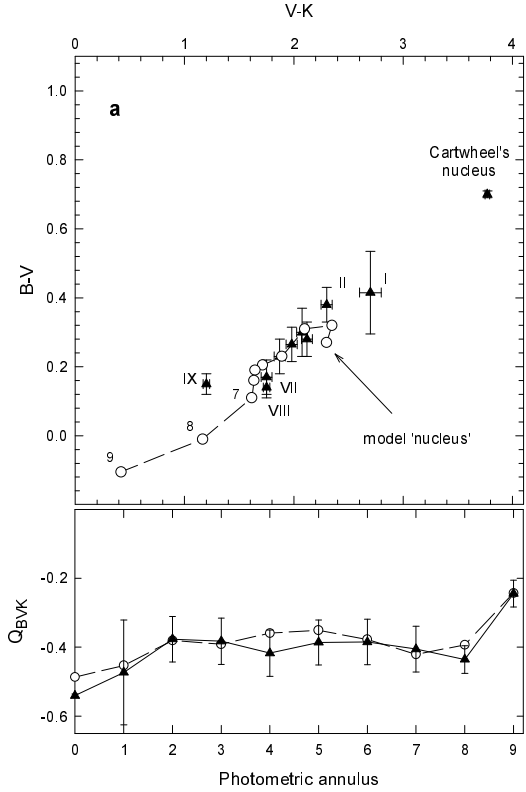}}
      \caption{({\bf a}) The $B-V/V-K$ color-color diagram, and ({\bf b}) 
the $Q_{\rm BVK}$ radial profiles. All parameters are the same as in Fig.~5, 
but with an additional 10-Myr-old burst of star formation in the Cartwheel's nuclear region.}
         \label{Fig7}
   \end{figure}  
In the cloud-fluid models of ring galaxies, Appleton $\&$ Struck-Marcell (\cite{Appleton2}) 
predicted that the infall of gas behind the ring should lead to a strong compression of gas in 
the central parts. In case of the Cartwheel galaxy, a large-scale radial infall of HI was reported 
by Higdon (\cite{Higdon2}). Recent observation in the mid-infrared 
(Charmandaris et al. \cite{Charm}) and in $\mathrm{H{\alpha}}$ line 
(Amram et al. \cite{Amram}) have shown that the Cartwheel's inner ring and
the nucleus might have a low level star formation activity. 

In this section, we perform  numerical modelling of the $B-V/V-K$ colors and 
$Q_{\rm BVK}$ indices measured in the Cartwheel's inner ring and the nucleus (Marcum et al. \cite{Marcum}).
To do this, we assume that the star formation takes place not only along the perimeter of the expanding  
density wave, but also in the central region of the Cartwheel.
We set off an outwardly propagating density enhancement of amplitude 
$A(r)=50\; \exp(-r/15) \; M_{\odot}\;\rm  pc^{-2}$
and velocity $v=55\;\rm km\; s^{-1}$ and compute the luminosity of pre-collision and post-collision stars 
as in Sect.~\ref{purewave}.
Then we add the luminosity of stars born in the central region  and use the  total luminosity 
in each annulus to calculate the resulting surface brightness, color, and index profiles. 

While the inner ring and the nucleus are currently
HI poor with $\Sigma_{\rm HI} \le 0.3\; M_{\odot}\;\rm pc^{-2}$ (Higdon \cite{Higdon2}), 
they have a significant amount of molecular hydrogen as suggested by 
recent CO observations of Horellou et al. (\cite{Horellou}). 
We assume that molecular hydrogen in
the Cartwheel is   exponentially distributed in the central 9.5~kpc, with the peak surface density of 
35 $M_{\odot}\;\rm pc^{-2}$  in the nucleus and the scale length of  3~kpc 
(radius of the inner ring). The total mass of $\mathrm{H_2}$  is then close 
$1.6 \times 10^9\; M_{\odot}$, as estimated by Horrelou et al. (\cite{Horellou}) 
for the solar metallicity. Metallicity in the central part of the Cartwheel is unknown. 
We assume that star-forming gas in the central 9.5~kpc is of solar metallicity, 
as suggested by numerical modelling in Sect.~\ref{purewave}.
 The star formation law in the nucleus and the inner ring is assumed to follow 
the classic Schmidt behavior:
\begin{equation}
SFR= a \; \Sigma_{\mathrm{H_2}}^{1.5},
\label{Schmidt}
\end{equation}
with the rate of star formation $a$ defined as in Eq. (\ref{eq1}). 

We use two conventional models for star formation: a constant star formation and a short starburst.
In both models star formation proceeds until $1.6 \times 10^8 \; M_{\odot}$ of stars is formed,
which is $10\%$ of the initial mass of gas. In the short starburst model we increase
the rate of star formation $a$ to shorten the star formation event.

{\it Constant star formation in the nuclear region for the last 60~Myr.} 
Assumption of the constant star formation activity in the Cartwheel's central region allows
us to reach a better agreement between the model and observed 
$B-V/V-K$ colors and $Q_{\rm BVK}$ indices of the nucleus and the inner ring. 
The model $Q_{\mathrm{BVK}}$ index of the inner ring  lies within 
the observed limits, which is not the case in Fig.~5d. Now, however, the 
inner ring+nucleus and the outer ring have comparable $\mathrm{H\alpha}$ fluxes.
This is problematic since observations  indicate that the Cartwheel lacks  
strong $\mathrm{H{\alpha}}$ emission from the nucleus and the inner ring, 
while it has a large $\mathrm{H{\alpha}}$ luminosity in the outer ring 
(Higdon \cite{Higdon1}, Amram et al. \cite{Amram}). 
We assume that the lack of  strong $\mathrm{H\alpha}$ 
emission  from the nucleus and the inner ring   is the result of a burst-like nature 
of star formation in the nuclear region.
If the active star formation in the nucleus+inner ring had ceased  about ten 
million years ago, then the observed $\mathrm{H\alpha}$ flux would have 
fallen considerably at present. 

{\it Short starburst in the nuclear region.} Infall of gas into central parts of a galaxy is 
known to trigger starbursts. We assume that  the large-scale radial infall of gas, 
reported by Higdon (\cite{Higdon2}), has triggered a short starburst in the 
Cartwheel's nucleus and the inner ring ten million years ago.
The resulting model $B-V/V-K$ radial color gradients and $Q_{\rm BVK}$ 
radial profiles are illustrated with the open circles in Figs.~7a and 7b respectively, 
while the filled triangles with error bars indicate the corresponding radial profiles observed in 
the Cartwheel galaxy. The sequence of $Q_{\mathrm{BVK}}$ indices observed in the galaxy is
well reproduced, as shown by the open circles in Fig.~7b.
Now, the model peak value of  $\mathrm{H\alpha}$ surface brightness  at the outer ring 
is 60 times larger than the model peak value at the nucleus, which is close to what is 
observed in the Cartwheel galaxy (Amram et al. \cite{Amram}).

The difference of the model and Cartwheel's $B-V/V-K$ colors of the inner ring (annulus I) 
and the nucleus, which is seen in Fig.~7a, can be
explained in terms of internal extinction using the same approach as in Sect.~\ref{purewave}. 
 Implied by this difference, color excess $E(B-V)$ of the Cartwheel's nucleus 
is $0^{\mathrm{m}}.43$, which is equivalent to 
$\Sigma_{\mathrm{HI+H_2}}$ of 19~$M_{\odot} \;\rm pc^{-2}$ 
(Bohlin et al. \cite{Bohlin}). If molecular hydrogen is exponentially distributed in the central 9.5~kpc,
then this value of $\Sigma_{\mathrm{HI+H_2}}$ might be below the detection upper limits of molecular hydrogen in the Cartwheel's nucleus. 
Hence, the internal extinction can be responsible for the difference between
the model and observed $B-V/V-K$ colors of the nucleus shown in  Fig.~7a. 
This conclusion is also true for  the Cartwheel's inner ring, 
where $\Sigma_{\mathrm{HI+H_2}}=4.5\; M_{\odot}\;\rm pc^{-2}$ is inferred 
using $E(B-V)=0^{\mathrm{m}}.1$.
Now, the young post-collision stellar populations and the old pre-collision stellar 
populations contribute equally to the total flux in the {\it V}- and {\it K}-bands at the smaller 
radii ($R \le 3~kpc$).

\section{Conclusions}

In this paper we model and analyse the $B-V/V-K$ radial color gradients observed in 
the Cartwheel ring galaxy. We use the $Q_{\rm BVK}$ combined color indices  
to analyse the Cartwheel's $B-V/V-K$ radial color gradients. Use of the $Q_{\rm BVK}$
indices minimises the uncertainties in the  observed   $B-V$ and $V-K$  colors introduced 
by dust extinction. We find that the $Q_{\rm BVK}$ radial profile observed in 
the Cartwheel's disk falls naturally into three sub-profiles, which correspond to
the nucleus+inner ring, the inter-ring region, and the outer ring of the galaxy.
Combined color indices are insensitive to dust extinction and can be directly understood 
in terms of stellar populations. This implies that the nucleus+inner ring, inter-ring region, 
and the outer ring of the Cartwheel might consist of different stellar populations.

To model the optical and near-infrared color properties of the Cartwheel galaxy, we use a toy model, 
which imitates the propagating star formation in the galactic disk. Namely, we assume that the 
star formation rate is proportional to a Gaussian density enhancement, expanding 
radially in a gaseous disk. We use the population synthesis to calculate the color properties 
of stellar populations formed by this density enhancement.

The results of color modelling strongly suggest that the pre-collision Cartwheel was a 
late-type spiral. The old stellar disk was embedded
in an extensive gaseous disk of sub-critical surface density.
Existence of the pre-collision stellar disk of low surface brightness
 $\mu_V^0=23.5 \;\rm mag \; arcsec^{-2}$, extending out to the current location of 
the Cartwheel's outer ring at 16~kpc ($R_0=5$~kpc), 
was first reported by Korchagin et al. (\cite{Korch}). However, we find that the 
pre-collision stellar disk might be smaller, extending out to $\approx 12$~kpc.
Properties of the pre-collision stellar disk are typical for the late-type Freeman disks 
($\mu_V^0=21.0 \;\rm mag \; arcsec^{-2}$ and $R_0=2$~kpc) rather than for  low surface 
brightness galaxies as found by Korchagin et al. (\cite{Korch}).
The pre-collision gaseous disk had a metallicity gradient ranging from $z=z_{\odot}/5$ 
at the outer parts to $z=z_{\odot}$ in the central parts.
Approximately 300 Myr ago, the Cartwheel galaxy collided face-on and near-centrally 
with a companion galaxy. This collision generated an expanding ring density wave, 
which triggered massive star formation along the wave's perimeter.  

The results of our modelling show that the young stellar populations, formed by the density wave, 
dominate in the outer ring. Thus, the  wave of star formation has passed the extent of 
the pre-collision stellar disk and is currently moving in the predominantly gaseous, 
low-metallicity disk at the radius of 16~kpc. 
In the region between the inner and the outer rings, however, 
the young stellar populations and the pre-collision stellar populations
contribute comparable fluxes in {\it K}-band. We find that the Cartwheel's $B-V/V-K$ colors
and $Q_{\rm BVK}$ indices of this inter-ring region cannot be successfully modelled without 
taking into consideration the relative input from the old, pre-collision stellar populations.
    
The observed colors of the nucleus+inner ring are most difficult to model. Neither
young stellar populations formed in the expanding density wave, 
nor their mixture with the old, pre-collision stellar populations can reproduce the 
$B-V$ and $V-K$ colors of the Cartwheel's nucleus+inner ring. 
We find that an additional 10-Myr-ago burst of star formation in the nuclear regions, 
along with the visual extinction of $A_V=1^{\mathrm{m}}.3$,  might be responsible for 
the peculiar colors of the Cartwheel's nucleus.
This is not totally unexpected, as some observational evidence 
points to the existence of a low-level star formation activity in the central regions (Amram et al. \cite{Amram}, 
Charmandaris et al. \cite{Charm}) and the large-scale radial infall of gas in the Cartwheel, 
reported by Higdon (\cite{Higdon2}), might have triggered bursts of star formation in the past.

\begin{acknowledgements}
The authors thank two anonymous referees for useful comments.
The authors are grateful to Dr. Y.D. Mayya for providing his population synthesis program. 
Our special thanks are to Dr. V. Korchagin for continuous encouragement. We also thank Dr. de Jong
for providing his observational data.
\end{acknowledgements}

\end{document}